\documentclass[twocolumn, pra, amssymb,superscriptaddress, aps,preprintnumbers,amsmath,floatfix]{revtex4}

\usepackage{amsmath}
\usepackage{amssymb}
\usepackage{color,soul}
\usepackage{graphicx}  
\usepackage{mathrsfs}
\usepackage{datetime}
\usepackage[hidelinks]{hyperref}
\usepackage{multirow}
\usepackage{isotope}

\usepackage[dvipsnames]{xcolor}

\usepackage{pdfpages}

\begin{document}

\title{Collective effects in $^{229}$Th-doped crystals}

\author{Brenden S. Nickerson}
\affiliation{Max-Planck-Institut f\"ur Kernphysik, Saupfercheckweg 1, D-69117 Heidelberg, Germany}

\author{Wen-Te Liao}
\affiliation{Department of Physics, National Central University, Taoyuan City 32001, Taiwan}

\author{Adriana P\'alffy}
\email{Palffy@mpi-hd.mpg.de}
\affiliation{Max-Planck-Institut f\"ur Kernphysik, Saupfercheckweg 1, D-69117 Heidelberg, Germany}


\date{Thursday 17$^{\text{th}}$ January, 2019}

\begin{abstract}

Vacuum-ultraviolet-transparent crystals have been proposed as host lattice for the coherent driving of the unusually low-lying isomer excitation in $^{229}$Th for metrology and quantum optics applications. Here the possible collective effects occurring for the coherent pulse propagation in the crystal system are investigated theoretically. We consider the effect of possible doping sites, quantization axis orientation and pulse configurations on the scattered light intensity and signatures of nuclear excitation. Our results show that for narrow-pulse driving, the rather complicated quadrupole splitting of the level scheme is significantly simplified. Furthermore, we investigate complex driving schemes with a combination of pulsed fields and investigate the occurring interference process. Our theoretical results support experimental attempts for first direct driving of the nuclear transition with coherent light.

\end{abstract}

\maketitle

\section{Introduction}
Today's global primary and secondary time standards are based on coherent light driving atomic transitions. However, efforts are underway to extend the clock physical systems to atomic nuclei. This development is based on a unique nuclear transition in the vacuum ultraviolet (VUV) range in the actinide nucleus $^{229}$Th \cite{Peik_Clock_2015}. The first excited state of $^{229}$Th is an isomer, i.e., a long-lived excited state, and lies at only 7.8 eV, being in the range of VUV lasers. Advantageous features of this isomeric transition towards the development of a nuclear clock include the very small ratio of radiative width to transition energy $\Gamma_\gamma/E_m\approx 5\times 10^{-21}$ (based on the theoretical predictions for $\Gamma_\gamma$ in Ref.~\cite{Minkov_Palffy_PRL_2017}) and the isolation from external perturbations promising amazing stability  \cite{Peik_Clock_2003}. At present, efforts in the development of a nuclear frequency standard are centered around a more accurate determination of the isomer state energy.  The currently accepted value $7.8 \,\pm\, 0.5$ eV dates back to indirect gamma-spectroscopy measurements performed in 2007 \cite{Beck_78eV_2007, Beck_78eV_2007_corrected} with the only direct measurement of the excitation (however without providing information on the  corresponding energy) coming more recently \cite{Wense_Nature_2016, Seiferle_PRL_2017}.

Direct excitation of the isomeric state has proven difficult due to the exact feature that makes it so promising, its narrow linewidth. To this end, one of the experimental approaches makes use of Th-doped VUV-transparent crystals \cite{Rellergert_2010, Dessovic_2014}, which render possible fluorescence spectroscopy on solid-state samples with a high isomer density. The ability to address a large number of $^{229}$Th nuclei should lead to fluorescence rates sufficient for the use of broadband synchrotron light to directly measure the transition energy \cite{Rellergert_2010}. Due to their large band gap, crystals like CaF$_2$ or LiCaAlF$_6$ should be transparent in the region of the isomer energy \cite{Dessovic_2014}. In practice, several attempts of direct photoexcitation of the isomeric state around the 7.8 eV value with broadband light sources have been unsuccessful \cite{Jeet_PRL_2015,Yamaguchi2015, Stellmer2018}. The two major sources of background that might cover the nuclear spectroscopy signal, namely  VUV photoluminescence, caused by the probe light, and radioluminescence, caused by the radioactive decay of $^{229}$Th and its daughters, have been investigated in Ref. \cite{Stellmer2015}. 

Since  inhomogeneous broadening in the crystal lattice environment compromises the traditional clock interrogation schemes, fluorescence spectroscopy was presented as an alternative \cite{Kazakov_2012}. A significant suppression of the inhomogeneous broadening is expected as long as all nuclei experience the same crystal lattice environment and are confined to the Lamb-Dicke regime, i.e., the recoilless transitions regime \cite{Dicke53,Rellergert_2010}. However, theoretical work has  shown that these very conditions lead to coherent light propagation through the sample and enhanced transient fluorescence in the forward direction, with a speed up of the initial decay (homogeneous broadening) depending primarily on the sample optical thickness \cite{Liao_2012}. These collective effects are actually well known from resonant coherent light scattering in different parameter regimes such as nuclear forward scattering (NFS) of synchrotron radiation \cite{Hannon1999} driving M\"ossbauer nuclear transitions in the x-ray regime, or from the interaction of atomic systems with visible and infrared light \cite{Crisp1970,Hartmann1983,Rothenberg1984}. The coherent enhancement of the resonant scattering in the forward direction can be exploited for a more efficient excitation, but also in combination with additional electromagnetic fields to provide a specific signature of the nuclear excitation. First proposals in this direction have been discussed in Refs.~\cite{Liao_2012,Das2013,Liao2017}.

In this paper we extend the study of collective effects for the  $^{229}$Th nuclear clock transition in VUV-transparent crystals and investigate theoretically several excitation schemes involving one or two VUV laser fields and a combination of VUV laser field and external magnetic field. We go beyond the previous results in Refs.~\cite{Liao_2012,Das2013,Liao2017} to show that each excitation scheme requires analysis of the crystal structure and dopant orientation, which give information not only on the hyperfine structure of the levels to be driven, but also on the orientation of the possible quantization axes. The hyperfine structure determines the required energies to drive transitions as well as the angular momentum selection rules in the frame of each individual Th nucleus. Knowledge of the quantization axis, which can be different between dopant sites, determines which transitions will be driven in the bulk sample by a defined laser-pulse polarization in the laboratory frame. We identify setups for the excitation of the nuclear clock transition in $^{229}$Th and the correct interpretation of the experimental fluorescence spectra. We also investigate the role of phase relations between VUV laser pulses for efficient nuclear excitation by a pulse train. Our results show that the time interval between pulses, detuning, and phase difference play a critical role for the scattered spectra. A setup comprising of two crystal samples, one of which under the action of a moderate external magnetic field, is shown to provide the desired nuclear excitation signature due to interference effects and a clear signal also when using VUV pulse trains.

The paper is structured as follows.  In Sec.~\ref{Th_model} the most relevant features of coherent excitation of $^{229}$Th in crystal environments along with the importance of the quantization axis are introduced. Our analysis allows for the simplification of the multi-level systems into two and three-level systems  showing both the enhancement of the radiative decay as well as the quantum beat signature in the case of the three-level system. The theoretical quantum optics model based on the Maxwell-Bloch equations for the pulse propagation through the $^{229}$Th:CaF$_2$ crystal is discussed in Sec.~\ref{MBE}.  Numerical results for excitation schemes using one or two VUV pulses in different configurations are presented in Secs.~\ref{numres1} and \ref{numres2}. Nuclear excitation with a pulse train is investigated along with the interplay between phase relation, detuning and pulse spacing in Sec.~\ref{seqpul}. Finally, Sec.~\ref{2samples} considers a two-crystal setup with a static magnetic field. Section \ref{Conclusions} summarizes our conclusions, followed by Appendixes outlining important details of crystal structure including dopant orientation, quantization axis, and state mixing.

\section{A model system: $^{229}\text{Th:CaF}_2$ \label{Th_model}}
\begin{figure}[htb!]
\centering
\includegraphics[scale=0.5]{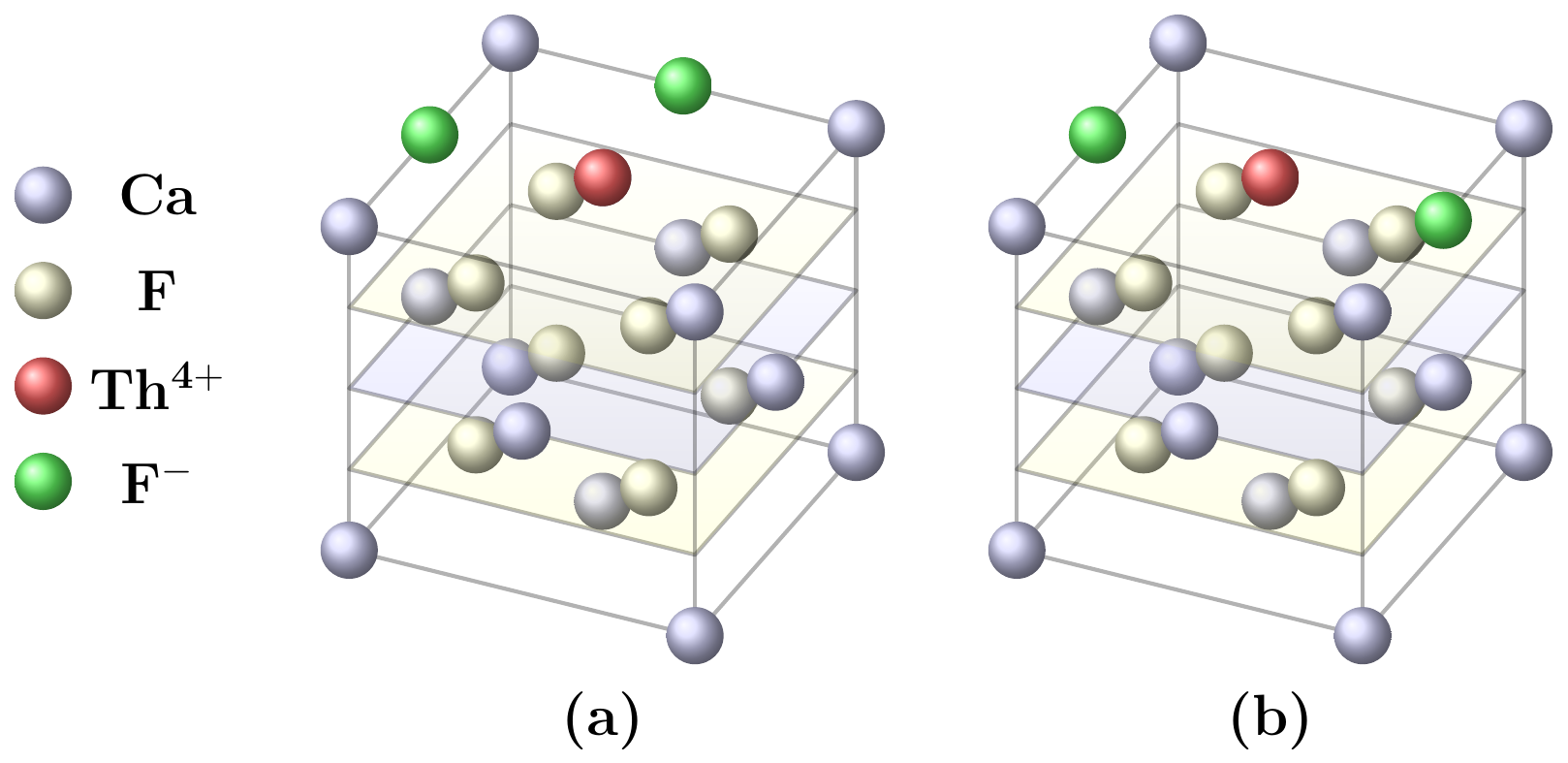}
\caption{Th:CaF$_2$ structure, dopant orientation with (a) $90^\circ$ and (b) $180^\circ$ fluoride interstitials.}
\label{fig:CaF290180_2}
\end{figure}

We consider in the following the case of $^{229}$Th nuclei doped in  CaF$_2$. The $^{229m}$Th isomer level is believed to lie within the band gap of CaF$_2$ measured in the range of $11-12$ eV \cite{Rubloff1972, Barth1990, Tsujibayashi2002}, rendering the crystal transparent in the  energy range of the nuclear transition. The CaF$_2$ crystal displays a cubic lattice structure. Thorium doped in the crystal has charge state $4+$ and replaces one of the calcium ions introducing two more interstitial fluorine ions for charge compensation. A density-functional study \cite{Dessovic_2014} shows that there exist preferred doping configurations, among which the two with lowest energy are the cases when the two fluorine interstitial ions are in a $90^\circ$ and a $180^\circ$ configuration, as illustrated in Fig. \ref{fig:CaF290180_2}.

As a result of the CaF$_2$ crystal environment, the doped $^{229}$Th nuclei experience quadrupole level splitting \cite{Tkalya_2011, Kazakov_2012, Dessovic_2014} according to the Hamiltonian
\begin{eqnarray}
&&\hat H_{E2} =\frac{eQV_{zz}}{4I(2I-1)}\left[3\hat{I}_z^2 - \hat I + \frac{\eta}{2}\left(\hat{I}_+^2+\hat{I}_-^2\right)\right],
\label{eqn:He2}
\end{eqnarray}
where $e$ is the electric charge, $Q_g=3.11\, \text{b}$ \cite{Campbell2011, 1402-4896-38-5-004} and $Q_e=1.8 \,\text{b}$ \cite{Thielking2018, Tkalya_2011} are the quadrupole moments of the ground and isomer state, respectively, where $\text{b} = 10^{-24}$ cm$^2$, $V_{zz}$ is the dominant component of the electric field gradient at the thorium nuclei, $I_g=5/2$ and $I_e=3/2$ are the nuclear spin angular momenta of the ground or isomer state, respectively, and $I_z=m$ its projection on the quantization axis. Furthermore, $\hat I$ and $\hat I_z$ are the angular momentum and projection operators with raising and lowering operators $\hat I_+$ and $\hat I_-$, respectively. Finally, $\eta = (V_{xx}-V_{yy})/V_{zz}$ is the asymmetry parameter of the electric-field gradient. 

The orientation of the quantization axis ($q$ axis) plays an important role for the NFS modeling. The $q$ axis along with the polarization of the exciting field determine the allowed transitions based on angular momentum selection rules. Dopant nuclei with aligned $q$ axes will undergo the same transitions when exposed to fields of like polarization, however, the same can in general not be said for nuclei with misaligned $q$ axes. The use of polarized fields to selectively drive transitions is therefore only possible if the majority of the dopant nuclei share a $q$ axis, or if misaligned nuclei do not contribute to the signal. Thus it is compulsory that one considers the impact of the crystal structure and $q$ axes at the different Th dopant sites for a reliable modeling of the scattering. However, certain simplifications can be made when considering driving by a narrow-band VUV laser as shown in Appendix \ref{AppA}. 

To interpret this complex system, we present in the following a number of assumptions concerning the sample and field in order to demonstrate their consequences on the scattering spectra. \\
{\bf (i)} To simplify the initial discussion we consider the second most likely case of $180^\circ$ fluoride interstitials where $\eta=0$ (no state mixing) and $V_{zz}=-296.7$ V\AA$^{-2}$.  As discussed in Appendix \ref{AppB}, the state mixing can safely be neglected also for the lowest energy case with $90^\circ$ fluoride interstitials and $\eta=0.48$ and all methods outlined here will still apply. \\
{\bf (ii)} We consider the case of coherent driving with a narrow-band laser. When driving transitions near resonance, our calculations in Appendix \ref{AppA} show that one can reduce the Th:CaF$_2$ 10-level scheme to a simpler effective system with uniform quantization axis. This is justified because the detuning in energy to the transition of interest $E_\Delta$ is less than the energy width of the excitation pulse $E_p$, where $E_p$ is less than the quadrupole level splitting $E_Q$ and does not overlap with multiple hyperfine levels, $E_\Delta<E_p<E_Q$. In such a case many levels can be neglected in systems with mismatched $q$ axis, because even though selection rules are satisfied, they are far out of resonance in comparison to the transitions of interest. In this limit, plots of relative intensity will show correct functional behavior but will have to be scaled by factors relating the fraction of population taking part in the transitions. \\
{\bf (iii)} We assume that crystal cooling will reduce the ground-state population to only the lowest hyperfine sublevels.

\begin{figure*}[htb!]
\centering
\includegraphics[scale=1]{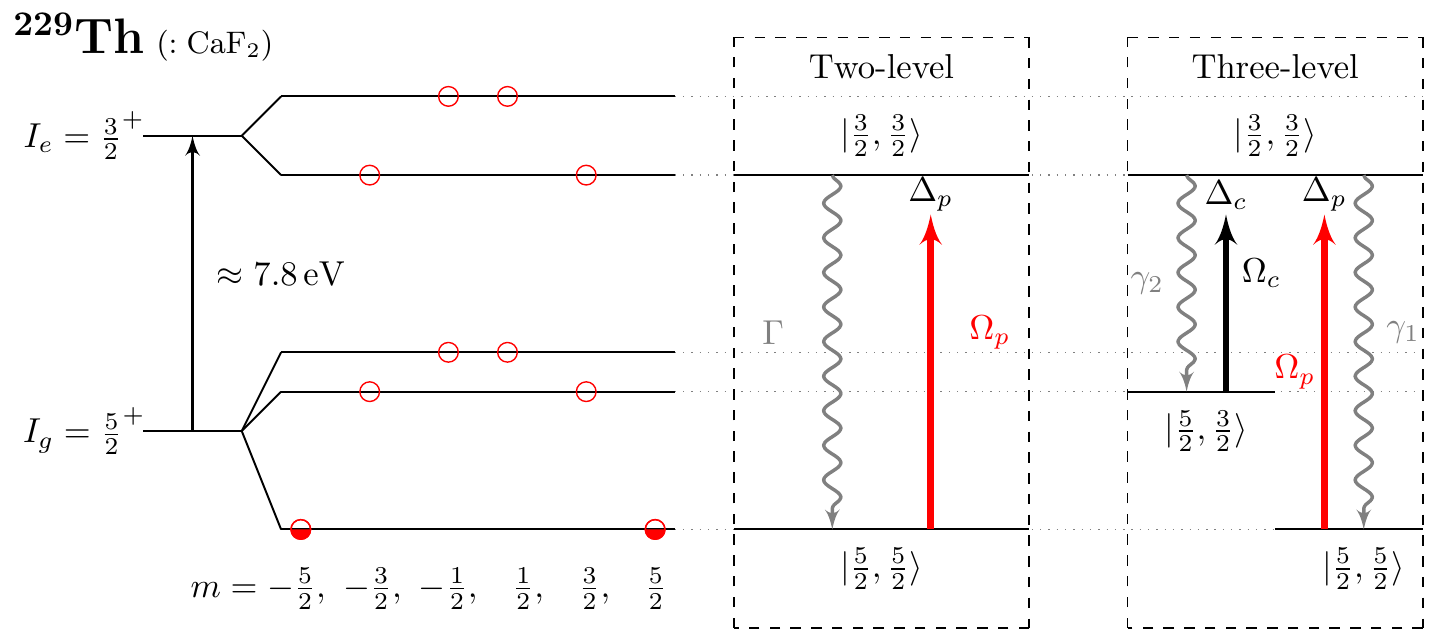}
\caption{Nuclear quadrupole splitting level scheme $^{229}$Th:CaF$_2$ (not to scale) for the $180^\circ$ fluoride interstitials doping configuration. Both ground state and isomeric state have positive parity. The possible projections of the nuclear spin angular momentum on the quantization axis are denoted by $m$. Assuming that only the two lower ground state hyperfine levels are populated, probe ($\Omega_p$) and couple ($\Omega_c$) VUV  laser pulses can couple to a two- and three level schemes depicted in the panels on the right-hand side. $\Delta_{p/c}$ stands for the corresponding detunings of the probe and couple laser pulses. }
\label{fig:thorium}
\end{figure*}
For the $180^\circ$ fluoride interstitials doping orientation, the hyperfine-split nuclear level scheme is illustrated in Fig.~\ref{fig:thorium}. Considering the positive parity of both the ground and isomeric states, selection rules require that the nuclear transition is of magnetic dipole ($M1$) character. Weak multipole mixing of an electric quadrupole ($E2$) channel is possible which can be safely disregarded when considering the radiative excitation or decay of the nucleus \cite{PavloE2}. The hyperfine splitting that occurs allows for several excitation schemes to be investigated such as a two-level system driven by one VUV laser field and a three-level system driven by two VUV laser fields illustrated in Fig.~\ref{fig:thorium}, addressed previously in Ref.~\cite{Liao_2012}.  Concentrating on the $|\frac{5}{2},\frac{5}{2} \rangle$ level as the initial state [see assumption (iii)] we construct the two- and three-level systems by choosing the appropriate driving lasers' orientation and polarization. Being an $M1$ transition we will refer to the polarization vector of the magnetic component of the exciting laser. First, a left-handed circularly polarized probe pulse $\Omega_p$ moving parallel to the quantization axis can be used to excite the isomeric $\Delta m=-1$ transition, $|\frac{5}{2},\frac{5}{2} \rangle \leftrightarrow |\frac{3}{2},\frac{3}{2} \rangle$. Secondly, for the three-level system, a linearly polarized continuous-wave (cw) couple laser $\Omega_c$ polarized parallel to the quantization axis and moving perpendicular to it drives the $\Delta m = 0$ transition, $|\frac{5}{2},\frac{3}{2} \rangle \leftrightarrow |\frac{3}{2},\frac{3}{2} \rangle$.

The dynamics of the system can be described using quantum optics methods. We employ  the Maxwell-Bloch equations \cite{Scully_QO_1999} to describe the nuclear population dynamics coupled to the dynamical coherent pulse propagation through the crystal sample. In the following we sketch the formalism for the more general case of two-field coupling to a three-level nuclear system.

\section{Maxwell-Bloch equations \label{MBE}}
The nuclear wave function of the three-level system in question can be written as $|\psi\rangle = A_1(t)|\frac{5}{2},\frac{5}{2}\rangle+A_2(t)|\frac{5}{2},\frac{3}{2}\rangle+A_3(t)|\frac{3}{2},\frac{3}{2}\rangle$ considering in the following the notation $|\frac{5}{2},\frac{5}{2}\rangle=|1\rangle$, $|\frac{5}{2},\frac{3}{2}\rangle=|2\rangle$, and $|\frac{3}{2},\frac{3}{2}\rangle=|3\rangle$. The density matrix is constructed via $\hat \rho = |\psi\rangle \langle\psi|$. We separate the Hamiltonian into two parts: the unperturbed Hamiltonian $\hat{H}_0$ of the nuclear system and the interaction part describing the laser-nucleus interaction $\hat{H}_{\rm int}$. The interaction Hamiltonian can be written in the most general form as 
\begin{equation}
\hat{H}_{\rm int}=-\frac{1}{c}\int d^3r\,\vec{j}(\vec{r},t)\cdot\vec{A}(\vec{r},t)\, ,
\end{equation}
where $\vec{j}(\vec{r},t)$ is the nuclear charge current and $\vec{A}(\vec{r},t)$ the vector potential of the laser field. Typically the interaction Hamiltonian can be expanded into nuclear multipole moments according to the driven transition, in our case, the magnetic dipole  multipole \cite{Ring1980}. We transform the Hamiltonian  into the interaction picture via the unitary transformation 
\begin{equation}
		\hat{U} =
 			\begin{pmatrix}
  			0 & 0 & 0  \\
   			0 & e^{-it (\nu_p-\nu_c)}  & 0 \\
  			0 & 0 & e^{-it \nu_p} 
 			\end{pmatrix},
\end{equation}
where $\nu_p$ and $\nu_c$ are the frequency of the probe and couple fields respectively. This results in 
the Hamiltonian $\hat{\tilde H} = i\hbar \partial_t \hat{U}^\dagger \hat{U} + \hat{U}^\dagger \hat{H}\hat{U}$ and density matrix $\hat{\tilde \rho} = \hat{U}^\dagger \hat{\rho}\hat{U}$. Making the rotating wave approximation (RWA) \cite{Scully_QO_1999} the resulting Hamiltonian of the three-level system is  
	\begin{equation}
		\hat{\tilde{H}} \overset{{R.W.A.}}{\approx} - \frac{\hbar}{2} 
 			\begin{pmatrix}
  			0 & 0 & C_{31}\Omega_p^*  \\
   			0 & -2(\Delta_p-\Delta_c) & C_{32}\Omega_c^*  \\
  			C_{31}\Omega_p & C_{32}\Omega_c & -2\Delta_p 
 			\end{pmatrix},
 	\end{equation}
where the matrix element of the interaction Hamiltonian $\Omega_{ij} = 2|\langle i |\hat{H}_{\rm int} |j \rangle|/\hbar $ with $i,j\in\{1,2,3\}$ is also known as the Rabi frequency of the $|j\rangle \rightarrow |i\rangle$ transition, with $\hbar$ the reduced Planck constant. Furthermore, $(C_{31}, C_{32}) = (\sqrt{2/3}, -2/\sqrt{15})$ are the Clebsch-Gordon coefficients and $\Delta_{p/c}$ are the detunings of the fields to their respective transitions.

Spontaneous decay and decoherence processes are included via the relaxation matrix
 	\begin{eqnarray}
		\hat{\tilde{\rho}}_r =  
 			\begin{pmatrix}
  			\Gamma C_{31}^2\rho_{33} & -\gamma_{12}^c\tilde{\rho}_{12} & -(\gamma_{13}^c +\frac{\Gamma}{2})\tilde{\rho}_{13}  \\
   			-\gamma_{21}^c\tilde{\rho}_{21} & \Gamma C_{32}^2\rho_{33} &-(\gamma_{23}^c +\frac{\Gamma}{2})\tilde{\rho}_{23}  \\
  			-(\gamma_{31}^c +\frac{\Gamma}{2})\tilde{\rho}_{31} &-(\gamma_{32}^c +\frac{\Gamma}{2})\tilde{\rho}_{32} & -\Gamma \rho_{33}
 			\end{pmatrix},
 	\end{eqnarray}	
where decoherence rates due to spin relaxation are for the $180^\circ$ doping configuration $(\gamma_{31}^c, \gamma_{32}^c, \gamma_{21}^c) = 2\pi\times(251, 108, 30)\, \text{Hz}$ \cite{Kazakov_2012}.

Finally, the Maxwell-Bloch equations describing the coherent pulse propagation of the probe field $\Omega_p$ are given by
\begin{eqnarray}
\partial_t \hat{\tilde \rho} = \frac{1}{i\hbar} [\hat{\tilde H},\hat{\tilde \rho}] +\hat{\tilde \rho}_r,
\label{eqn:bloch}
\end{eqnarray}
\begin{eqnarray}
 	\frac{1}{c}\partial_t\Omega_p+\partial_z\Omega_p = i \eta \,C_{31}\,\rho_{31} .
 \label{eqn:field}
\end{eqnarray}
Here the field equation for the  coupling laser, $\frac{1}{c}\partial_t\Omega_c+\partial_z\Omega_c = i \eta \,C_{32}\,\rho_{32}$, can be neglected as the high intensity of the coupling pulse is negligibly affected by the sample.
The parameter $\eta = 2 \xi \Gamma / L$ contains the dimensionless effective thickness $\xi = N\sigma L /4$ \cite{shvy_1998, chen_2007}, which is determined by the number density of nuclei $N$, the sample thickness $L$, and the resonant cross section \cite{Kong2014}
\begin{equation}
\sigma = \frac{2 \pi}{k^2} \frac{2I_e+1}{2I_g+1}\frac{\Gamma_\gamma}{\Gamma}\, .
\end{equation}
The origin of $\eta$ is the macroscopic current density  for a single nuclear resonance obtained by summing over all nuclei participating in the coherent nuclear scattering \cite{Shvydko1999,Kong2014}. Thus the factor $\eta$ in Eq.~(\ref{eqn:field}) is describing 
 the accumulation effect due to nuclei at different positions of the target during the pulse propagation for the forward mode. 
 We assume that no other decay channels are allowed in the crystal environment other than radiative decay; therefore, the radiative decay rate is equal to the total decay rate of the isomeric state $\Gamma_\gamma = \Gamma$. The resonant cross section is thus $\sigma = \lambda^2/(3\pi)$. Taking the currently accepted isomeric transition energy of $7.8$ eV results in $\sigma \approx 3\times 10^{-11}\, \text{cm}^2$.

Solving the Maxwell-Bloch equations, one can plot the NFS time spectrum, where the intensity $I\propto |\Omega_{p}|^2$ for the probe field. We note here that the Maxwell-Bloch equations are an equivalent approach to the iterative field equation method commonly used in NFS to describe the coherent scattering and so-called collective effects in coherent light propagation \cite{Shvydko1999}. In the field of NFS, collective effects refer to  the formation of a delocalized excitation extended over a large part of the sample. 
 This delocalized excitation, which typically does not contain more than one single excited nucleus, is also known as ``nuclear exciton'' and resembles a Dicke state \cite{Dicke54}. The formation of the exciton requires the indiscernibility of the possible scattering paths, i.e., recoilless transitions, no spin flips or internal conversion. This is the case of coherent scattering when the nuclei return to their initial state, such that the scattering path and the number of occurred events are unknown. The decay of the exciton  occurs via a complicated temporal structure known as the dynamical beat, which presents a speed-up decay at short times immediately after the excitation and additional damping and oscillations at later times. The dynamical beat can be very different from the natural decay of a single nucleus. Its origin is related to the process of coherent multiple scattering of a single resonant photon in the sample. 

In comparison with typical atomic physics superradiance, the collective effects in nuclear ensembles have to take into account two peculiarities. First, the condition that the wavelength is much larger than the internuclear distance often does not hold, since typically nuclear transitions are in the range of tens or hundreds of keV energy. $^{229}$Th is an exception, with the wavelength at $\sim$160 nm. Second, the induced nuclear excitation is very small, i.e., typically one or very few nuclei are excited. This is in contrast  to typical atomic superradiance effects which become most pronounced when approximately half of the atoms are excited. The different excitation phases in the structure of the nuclear exciton add up constructively only in the forward direction or at the Bragg angle for the case of scattering off crystal samples. In the forward direction, this leads to the  appearance of the complicated dynamical beat. A quite comprehensive review on the topic of nuclear excitons and collective effects in nuclear condensed-matter physics is given in Ref.~\cite{Hannon1999}. In our numerical results, the dynamical beat feature is not obvious when looking at the graphs because the decay due to the decoherence rates is the dominant component.

\section{Unique signature \label{numres1}} 
The multilevel structure of $^{229}$Th:CaF$_2$ allows for the study of a variety of subsystems. Here we apply the Maxwell-Bloch formalism to the two- and three-level systems illustrated in Fig. \ref{fig:thorium}.

Equations \eqref{eqn:bloch} and \eqref{eqn:field} are solved with initial conditions corresponding to a Gaussian input probe pulse and a cw couple laser,
\begin{eqnarray}
	\rho_{ij}(z,0)&=&\delta_{i1}\delta_{j1},\\
	\Omega_p(z,0) &=& \Omega_{p0} e^{-(t_p/T)^2},\\
	\Omega_p(0,t) &=& \Omega_{p0} e^{-((t-t_p)/T)^2},\\
	\Omega_c(z,t) &=& \Omega_{c0},
\end{eqnarray}
where $T = 10$ $\mu$s controls the pulse width and $t_p = 50$ $\mu$s the pulse delay time. The peak amplitude
is given by \cite[p15]{WTL_2013}
\begin{eqnarray}
	\Omega_{p/c0} = \frac{4}{\hbar 3!!} \sqrt{\frac{2\pi I_0}{c\epsilon_0}(2 I_g+1) \mathbb{B}(M1)},
\end{eqnarray}
where $\mathbb{B}$ is the reduced transition probability for the nuclear $M1$ transition which has been evaluated theoretically to $B_{W}(M1;3/2^+ (7.8\,\text{eV})\rightarrow 5/2^+(0.0\,\text{eV})) \approx 0.7 \times 10^{-2}$ Weisskopf units, converts via $B(M1;i\rightarrow f) = B_{W} \times 1.790 \times \mu_N^2$ where $\mu_N = \frac{e\hbar}{2m_pc}$ \cite{Tkalya_2015, Minkov_Palffy_PRL_2017} and $m_p$ is the proton mass. $I_0$ is the intensity of the exciting lasers, which for the couple was chosen to be 2 kW/cm$^2$. The normalized NFS intensity, $I= |\Omega_{p}/\Omega_{p0}|^2$, is independent of the chosen probe intensity $I_{p0}$ provided no Rabi oscillations occur while the pulse is active, hence provided $\Omega_{p0}<1/T$.

The detuning is taken here as $\Delta = \Delta_c=\Delta_p$ \cite{Liao_2012}, along with $L=1$ cm and $N=10^{18}\,\text{cm}^{-3}$ \cite{Kazakov_2012} which give $\xi \approx 7 \times 10^6$. The radiative lifetime of the isomeric state $^{229m}$Th is in the region of several hours, $\tau \approx 1$ h [$B(M1) \approx 0.032$ Wu. \cite{Tkalya_2011}], $\tau \approx 6$ h \cite{Zhao_2012}, and $\tau \approx 4.7$ h [$B(M1) \approx 0.007$ Wu. \cite{Minkov_Palffy_PRL_2017}]. Here we consider the decay rate as $\Gamma_\gamma = 1/\tau \approx 1\times 10^{-4}\, \text{s}^{-1}$, which is consistent with recent measurements of the internal conversion rate of $^{229m}$Th \cite{Seiferle_PRL_2017}.

\begin{figure}[h!]
\centering
\includegraphics[scale=1]{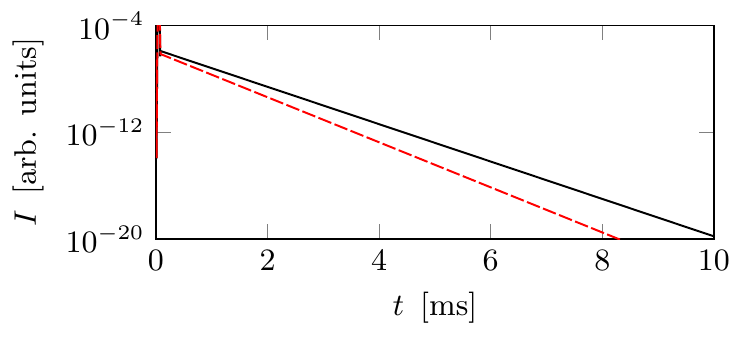}
\caption{Scattered intensity for the case of the two-level system for $\Delta = 0 \sim 10^8\Gamma$ (black solid line) and  $\Delta = 10^9 \Gamma$ (red dashed line). The excitation occurs for $\Delta < 10^{10} \Gamma \approx \hbar/E_p$. }
\label{fig:th2}
\end{figure}
The scattered intensity for driving the two-level system is illustrated  in  Fig.~\ref{fig:th2}, which shows the exponential decay of the isomeric state. The rate of decay of the NFS spectrum intensity is dependent not only on the radiative decay of the level population but also on the decay of the coherence in the system. As such the  exponential decay rate that can be observed in the calculated NFS spectrum is $\Gamma_F = \Gamma_{coh} + \xi \Gamma_\gamma + f(\Delta)$, where $f(\Delta)$ is a function of the laser detuning to the driven transitions where $f(0)=0$. The NFS intensity for this two-level system displays an exponential decay rate $\Gamma_F = 2\gamma^c_{31} + \Gamma + \xi C_{31}^2\Gamma +f(\Delta_p)$ where the largest contribution comes from the decay of the coherence $\Gamma_{coh} = 2\gamma_{31}^c + \Gamma$. Other than the exponential decay of the intensity, this two-level system does not provide any unique features that could help us differentiate it from other decay channels experimentally. 

The signature that we aim to create is that of quantum beats induced by interference. This can be typically induced by considering an effective V-type system. Here we consider two ways of constructing an effective V-type system: first, in this section, using a second laser to create Autler-Townes splitting of the excited state and second, in Sec.~\ref{2samples}, by exciting two crystals successively using the same laser pulse where one of the two crystals is in a static magnetic field.  

\begin{figure}[h!]
\centering
\includegraphics[scale=1]{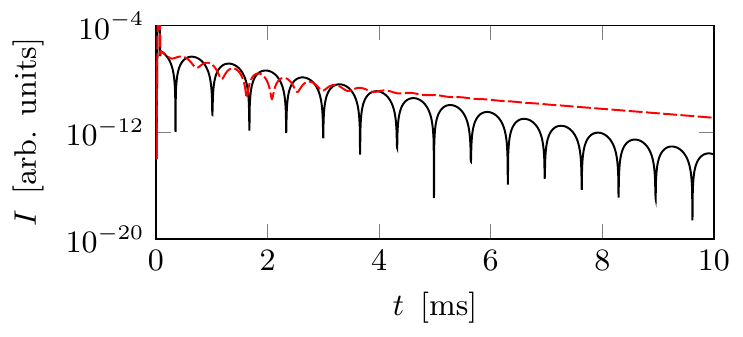}
\caption{ Scattered intensity for the case of the three-level system for  $\Delta = 0 \sim 10^6\Gamma$ (black solid line) and  $\Delta = 10^8 \Gamma$ (red dashed line). The excitation occurs for $\Delta < 10^{10} \Gamma \approx \hbar/E_p$.}
\label{fig:th3}
\end{figure}
When the strong couple laser is active, both the second and third levels experience Autler-Townes splitting whereby each level splits into two. For a small detuning $\Delta_c\ll\omega_{32}$, the energy separation of the splitting is $\hbar \Omega_{32} = \hbar |C_{32}| \Omega_c$, where each split level is displaced $\pm \hbar \Omega_{32}/2$ around the unsplit level energy \cite{Shore_CAE_1990, Boyd_NLO_2008}. The quantum beat in the NFS spectrum  as a result of the split third state, $| \frac{3}{2}, \frac{3}{2}\rangle$, decaying to the first, $| \frac{5}{2}, \frac{5}{2}\rangle$, is shown in Fig.~\ref{fig:th3}. The frequency of the quantum beat depends on the difference in energy of the two transitions \cite{Smirnov_1999} and hence the energy splitting of the third state. In this case, the frequency of the quantum beat is $f_{QB}= \Omega_{32}/2\pi$ and the minima occur at times $t^{min}_n= (n+\frac{1}{2})/f_{QB}+t_p$, where $n$ is an integer. Hence the larger the splitting, the smaller the time separation between minima. Note that the beat frequency in Fig.~2 of Ref.~\cite{Liao_2012} is in error by $\sqrt{2I_g+1}=\sqrt{6}$ due to this missing factor in the initial calculation of $\Omega_c$.

\section{Modified couple laser \texorpdfstring{$\Omega_c$}{O_c} \label{numres2}}
The experimental realization of a cw VUV laser is technically difficult. As such, the use of a pulsed coupling laser $\Omega_c$ would simplify the experimental implementation of the thorium three-level system shown above. In the following we consider the case of a three-level system driven by  pulsed probe and couple lasers for two couple pulse shapes.

\subsection{Square pulse}
\begin{figure}[h!]
\centering
\includegraphics[scale=1]{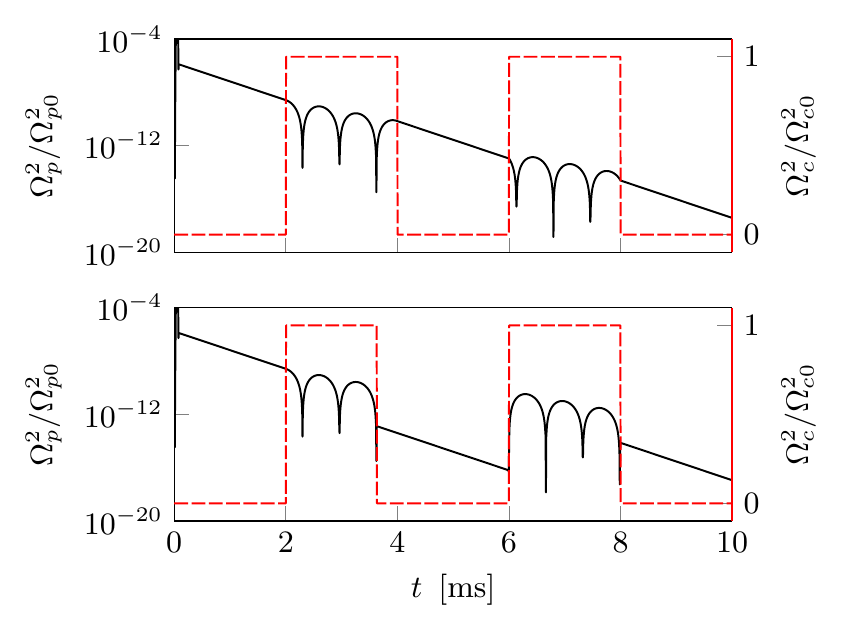}
\caption{NFS intensity of the three-level thorium system resulting from abruptly switching the couple laser on and off (black solid line, left axis) together with the relative intensity of the couple pulse (red dashed line, right axis).}
\label{fig:Oc_pulse1}
\end{figure}
Before considering a Gaussian pulse shape we examine the result of abruptly turning the cw laser on and off again after excitation by the probe pulse. For this we modify the initial conditions of the couple to be that of a square pulse 
\begin{eqnarray}
	\Omega_c(z,0) &=& 0\\
	\Omega_c(0,t) &=&  \begin{cases}
                    \Omega_{c0},  &   t_\uparrow< t < t_\downarrow \\
                    0, & \text{otherwise} 
                           \end{cases} 
\end{eqnarray}
where $(t_\uparrow,t_\downarrow)$ are the turn on and off times, respectively. As expected the beating only occurs while the couple laser is on, i.e. while the upper state is split; otherwise the intensity spectrum returns to that of the two-level system. Furthermore, the beating does not restart every time the couple is turned off and then on, but rather continues from where it previously ended. Similar to the ideas of storage via magnetic switching \cite{Liao_storage_2012}, if the couple is switched off at a minimum of the quantum beat, the beating can be revived with maximal intensity in the next on cycle as illustrated in  Fig.~\ref{fig:Oc_pulse1}.  

If we then consider many on-off cycles where the time spacing between the end of one cycle and the start of the next goes to zero, the NFS spectrum will approach that of a cw laser, i.e., Fig.~\ref{fig:th3}. As such we can envisage the use of a pulsed couple laser where the pulse spacing is far less than the time scale of the quantum beat, i.e., $\ll 2\pi/\Omega_{32}$.

\subsection{Gaussian pulse}
To convert $\Omega_c$ from cw to a Gaussian pulse laser we introduce a Gaussian pulse shape to the initial conditions
\begin{eqnarray}
	\Omega_c(z,0) &=& \Omega_{c0} e^{-(t_c/T_c)^2}\\
	\Omega_c(0,t) &=& \Omega_{c0} e^{-((t-t_c)/T_c)^2}
\end{eqnarray}
where $t_c$ is the delay of the pulse and $T_c$ is the half width of the pulse.  Similar to the case of the square pulse, the choice of $t_c$ and $T_c$ determine the time interval over which the quantum beat is visible. Provided there is no time delay between the couple and probe, $t_c = t_p$, and the width of the pulse $T_c\gg 10$ ms, the result is the same as shown in Fig.~\ref{fig:th3}. 
\begin{figure}[h!]
\centering
\includegraphics[scale=1]{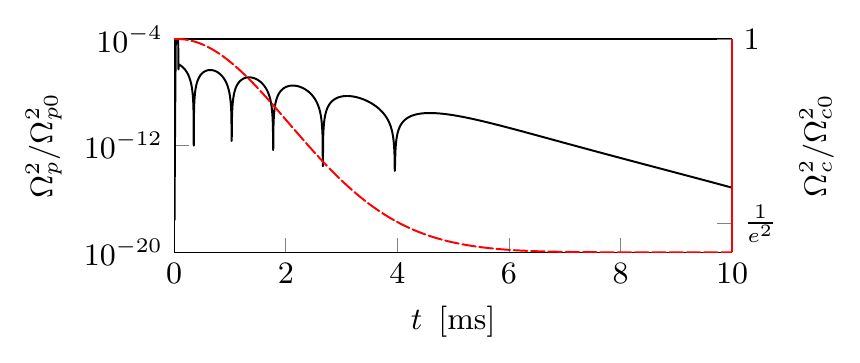}
\caption{NFS scattered intensity for the three-level thorium system with a Gaussian pulse couple laser (black solid line, left axis) together with the  couple pulse relative intensity (red dashed line, right axis) for  $t_c=t_p = 50$ $\mu$s and $T_c=4$ ms.}
\label{fig:Oc_pulse2}
\end{figure}

Our numerical results for the Gaussian couple pulse are illustrated in Fig.~\ref{fig:Oc_pulse2}.  Reducing the temporal width of the couple pulse will begin to erase the quantum beat in the region outside $T_c$. This is because, for a Gaussian pulse, the period of the induced quantum beat
\begin{eqnarray}
 	T_{QB} &=& \frac{2\pi}{|C_{32}|\Omega_{c0} e^{-((t-t_c)/T_c)^2}}
\end{eqnarray}
increases gradually as the pulse intensity diminishes. This is in contrast to $T_{QB}\rightarrow \infty$ in the case of the square-pulse couple-laser abruptly turning off.

This can be easily generalized to a pulse train with pulse spacing $\delta$
\begin{eqnarray}
\Omega_c &=& \Omega_{c0} \sum_{n=1}^N e^{-((t-t_c- n\delta)/T_c)^2}.
\end{eqnarray}
When both width and spacing of the Gaussian pulses become small compared to the time scale of the quantum beat at peak intensity, $(T_c,\delta) \ll 2\pi/(|C_{32}|\Omega_{c0})$, the resultant spectrum tends towards that of a cw laser.

\section{Train of probe pulses\label{seqpul}}
In experiment the intensity of the resonant pulse is usually weak, i.e., much fewer resonant photons per pulse than number of nuclei in the sample. As such, to generate the NFS signal, many pulses are used and the sum of the measured counts and time delays are used to build the final intensity spectrum. A variety of pulse shapes can be constructed from a single mode wave $E(t) = \cos(\nu t+\phi)$ by multiplying with the desired envelope; for simplicity we use a Gaussian $E(t) = e^{-((t-t_0)/T)^2}\cos(\nu t+\phi)$. Here we study the effect of introducing multiple excitation pulses varying both time delay and relative phase.

\subsection{Two-level system}
\subsubsection{$\Delta = 0$}

\begin{figure}[h!]
\centering
\includegraphics[scale=1]{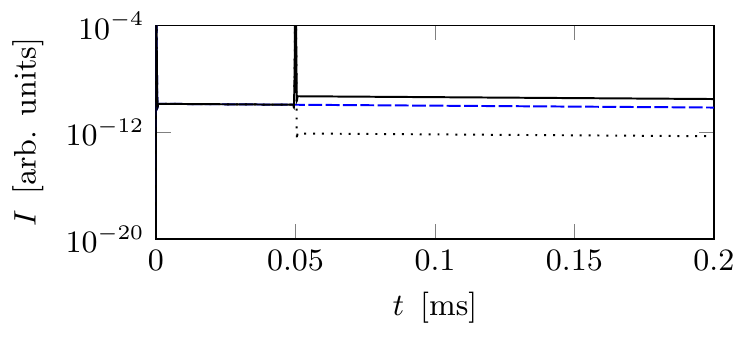}
\caption{Scattered intensity for the two-level thorium system with $\Delta = 0$, $t_0=50$ $\mu$s, and $T=0.1$ $\mu$s. (Blue dashed line) NFS intensity after the first pulse impacting at $t=0$. (Black solid line) With the additional excitation of a second pulse in phase with the first $\phi=0$. (Black dotted line) With the additional excitation of a second pulse with $\phi=\pi$.}
\label{fig:DG0_1}
\end{figure}

First, in order to identify the main features of the problem we consider for simplicity driving the two-level system seen in Fig.~\ref{fig:thorium} with no laser detuning $\Delta=0$ such that the exciting field has the same frequency as the transition. For further discussion we consider next to the total scattered intensity also the square of the real $\Re \{\Omega\}^2$ and imaginary $\Im \{\Omega\}^2$ components. In the case of zero detuning $(\Im \{\Omega\})^2 = 0$, and the NFS intensity is $I=|\Omega/\Omega_0|^2 = (\Re\{\Omega\}/\Omega_0)^2$. By changing the initial conditions of the calculation we can add a phase shift to the pulse
\begin{eqnarray}
\Omega(0,t) &=& \Omega_{0} e^{-(t/T)^2}e^{i\phi},
\end{eqnarray}
which for a single pulse has no effect on the resultant intensity spectrum. When introducing multiple excitation pulses the relative phase becomes important. Figure \ref{fig:DG0_1} shows the result of considering two pulses with a relative phase shift of $\phi = (0, \pi)$,
\begin{eqnarray}
\Omega(0,t) &=& \Omega_{0} \left(e^{-(t/T)^2}+e^{-((t-t_0)/T)^2}e^{i\phi}\right),
\label{eqn:second_pulse}
\end{eqnarray}
with a time delay $t_0=50$ $\mu$s and pulse width $T=0.1$ $\mu$s, where both incoming pulses have the same intensity. The incoming pulses arriving in phase to each other lead to constructive interference and an increased signal, while the pulses arriving in antiphase cancel each other and lead to a decreased signal in comparison with the case of a single incident pulse.

In more detail, this variation in NFS intensity is a manifestation of superposition where the intensity is proportional to the number of excited nuclei. The $m^{th}$ pulse excites $N_m(0)$ nuclei which have the option to decay back to the ground state $N_m(t) = N_m(0)\sqrt{e^{-\Gamma_F t}}$, but cannot be excited further in the two-level system. Provided there are enough nuclei in the ground state and all weak pulses $m$ have the same resonant intensity, then all $N_m(0)$ are equal. In such a case the total intensity of the emitted decay signal after a train of excitation pulses is 
\begin{eqnarray}
I(t) \propto \left|\sum_m N_m(t)e^{i\phi_m}\right|^2.
\end{eqnarray}
Considering just two pulses as before where $t_0$ is the delay time, the intensity $\{I(t)|t>t_0\}$ is
\begin{eqnarray}
I(t) &\propto& 	\left|N \sqrt{e^{-\Gamma_F t}}+N \sqrt{e^{-\Gamma_F (t-t_0)}}e^{i\phi}\right|^2\\
		&=&	N^2\left(e^{-\Gamma_F t}+e^{-\Gamma_F (t-t_0)}+2e^{-\Gamma_F (2t-t_0)/2}\cos\phi\right). \nonumber
\end{eqnarray}
The maximal (minimal) intensity is reached when there is no separation between the pulses and they add constructively $\phi=0$ (destructively $\phi=\pi$). Furthermore, it should be clear that, for long delay time, i.e., $t_0\gg1/\Gamma_F$, the initial excited population decays to the ground state and the intensity after the second pulse becomes once again independent of phase $I(t) \approx N^2e^{-\Gamma_F (t-t_0)}$.

\subsubsection{$\Delta \ne 0$}
\begin{figure}[h!]
\centering
\includegraphics[scale=1]{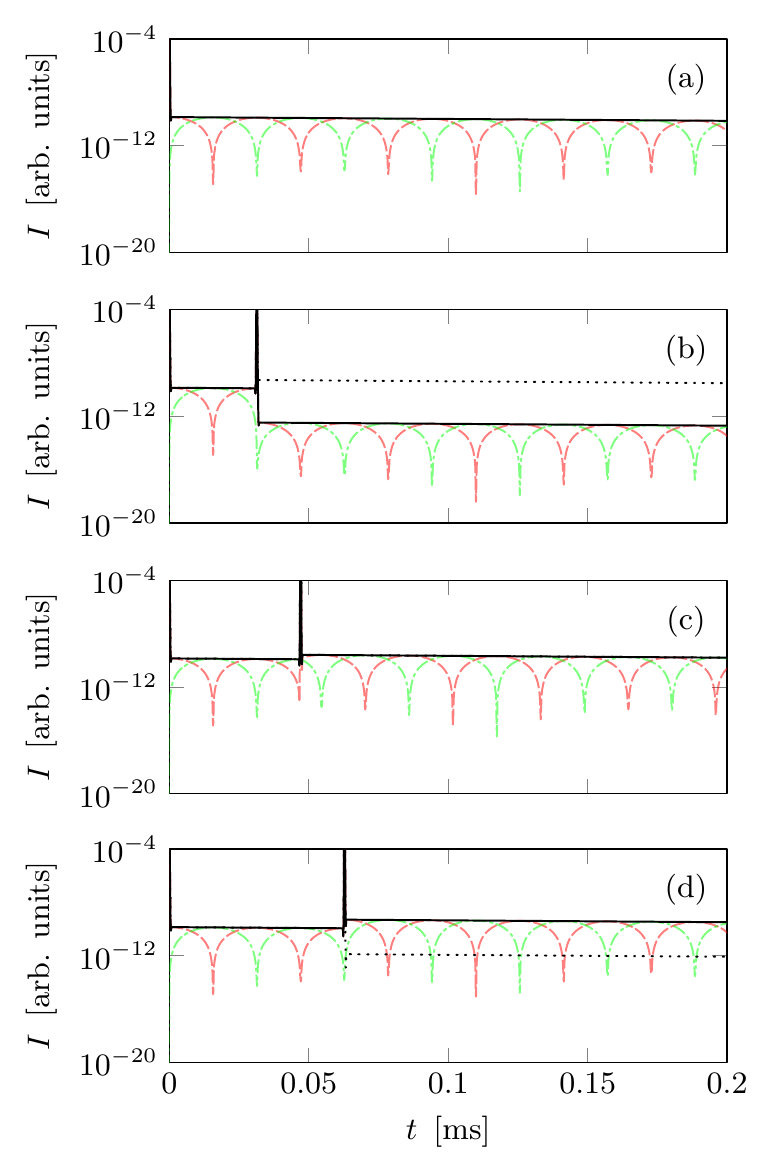}
\caption{ Scattered intensity for the two-level thorium system with $\Delta=10^9\Gamma$ and $T=0.1$ $\mu$s for (a)  a single pulse at $t=0$, (b) the first pulse at  $t=0$ followed by the second one at $t_0 = \pi/\Delta$, (c) the same but with delay between pulses $t_0 = 3\pi/2\Delta$, and (d) for delay between pulses $t_0 = 2\pi/\Delta$. (Black solid line) NFS intensity $|\Omega/\Omega_0|^2$ when using two pulses of the same phase. (Black dotted line) NFS intensity $|\Omega/\Omega_0|^2$ when using two pulses with phase shift $\pi$. (Red dashed line) $(\Re\{\Omega\}/\Omega_0)^2$. (Green dash-dotted line) $(\Im\{\Omega\}/\Omega_0)^2$. }
\label{fig:DG9_1}
\end{figure}

Next, we consider the same system as above but now driven with a detuning $\Delta=10^9\Gamma$. In the case $\Delta \ne 0$ both the real $\Re \{\Omega\}$ and imaginary $\Im \{\Omega\}$ components of the NFS intensity have a nonzero value. Figure \ref{fig:DG9_1}(a) shows the oscillation of the squared real (red dashed line) and imaginary (green dash-dotted line) components along with the NFS intensity (black solid line) as a result of a single pulse centered at $t=0$. The oscillation frequencies of $\Re \{\Omega\}$ and $\Im\{\Omega\}$ correspond to the detuning of the laser to the transition frequency,  $f_\Omega = \Delta/{2\pi}$. Alternatively for the intensity, due to the square of the amplitude, i.e., $\Re\{\Omega\}^2$ and $\Im\{\Omega\}^2$, the oscillation frequency is $f_{\Omega^2} = \Delta/{\pi}$.

Introducing additional pulses is done by modifying the initial conditions as shown earlier in Eq.~\eqref{eqn:second_pulse}. The results for $\phi = (0, \pi)$ with $t_0 = (\pi/\Delta, 3\pi/2\Delta, 2\pi/\Delta)$ are shown in Figs. \ref{fig:DG9_1}$(b)-$\ref{fig:DG9_1}(d). Clearly in the case of $\Delta \ne 0$, not only the relative phase $\phi$ but also the time separation $t_0$ are of critical importance. This is expressed in the intensity as
\begin{eqnarray}
I(t) &\propto& 	\left|N \sqrt{e^{-\Gamma_F t}}e^{i\Delta t}+N \sqrt{e^{-\Gamma_F (t-t_0)}}e^{i(\Delta(t-t_0)+\phi)}\right|^2 \nonumber \\
		&=&	N^2 \left(e^{-\Gamma_F t}+e^{-\Gamma_F (t-t_0)} \right. \nonumber \\
		&& \,\,\,\,\,\,\,\,\,\,\,\,\,\,\,\,\,\,\,\,\,\, \left. +2e^{-\Gamma_F (2t-t_0)/2}\cos(\Delta t_0 - \phi)\right).
\end{eqnarray}
For $\phi=0$, maximal (minimal) intensity happens for $t_0 = n\pi/\Delta$, where $n$ is an even (odd) integer. Therefore, by choosing the correct timing and phase we can use multiple pulses to cause increased excitation within the sample that will add constructively at the detector resulting in increased signal intensity. 

When searching for a transition with an unknown energy, the case of zero laser detuning is experimentally unlikely. Excitation will first be seen when driven with some nonzero detuning. However, to cause excitation we need to know the energy of the transition well enough to enforce $E_{\Delta i}<E_p$, i.e., the energy width of our pulse has to be wider than the detuning. 

To reliably increase the NFS decay intensity with a train of pulses, each successive pulse must cause a constructive excitation in the sample. For this to be the case the detuning of the laser to the excitation must be known. Because this is not the initially true in the search for the Th isomeric state, we must prevent the chance of destructive interference between the excitation caused by neighboring pulses. To this end, excitation pulses should be spaced out such that $t_0 \gg 1/\Gamma_F$. In this scenario, increasing the NFS intensity can only be accomplished by increasing the number of resonant photons in a single pulse.

If resonance is found in the two-level system, then additional excitation pulses can be introduced. The time delay between the additional excitation pulse can then be varied to find the peak intensity. By plotting the change in intensity as a function of pulse delay, the detuning could be more accurately determined. Repeating the process with a larger pulse train will serve to increase the accuracy of this determination.

\subsection{Three-level system}
 \begin{figure}[h!]
\centering
\includegraphics[scale=1]{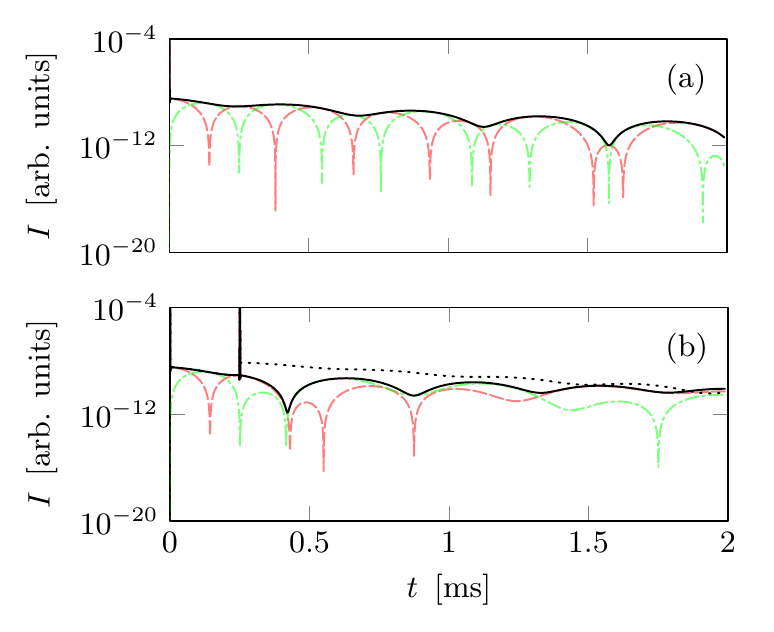}
\caption{Three-level thorium system driven by (a)  a single pulse at $t=0$ or (b) a single pulse at $t=0$ followed by a second identical pulse at $t=t_0$, which is chosen as the first minimum of $(\Im\{\Omega\}/\Omega_0)^2$.  (Black solid line) NFS intensity $|\Omega/\Omega_0|^2$ when using two pulses of the same phase. (Black dotted line) NFS intensity $|\Omega/\Omega_0|^2$ when using two pulses with phase shift $\pi$. (Red dashed line) $(\Re\{\Omega\}/\Omega_0)^2$. (Green dash-dotted line) $(\Im\{\Omega\}/\Omega_0)^2$. For the calculation we have used $\Delta=10^8\Gamma$, and $T=0.5$ $\mu$s.}
\label{fig:3s-DG8}
\end{figure}
The same train of probe pulses can also be applied to drive the three-level thorium system. Analogous to the two-level system, when there is zero detuning the probe pulses of the same phase will always add constructively to the excitation in the sample. However, for nonzero detuning, as shown in the two-level case, both the phase and delay time between pulses play a critical role in the intensity of the NFS spectrum. It is clear here, however, that the complication of this functionality grows quickly with the number of energy levels participating in the signal's generation, as illustrated in Fig.~\ref{fig:3s-DG8}, which shows the calculated spectrum for the three-level system. Our earlier approximation of the  $(\Re\{\Omega_p\})^2$ and $(\Im\{\Omega_p\})^2$ oscillation frequency as $f_{\Omega^2} = \Delta_p/{\pi}$ is inapplicable here. As a result, multiple pulses spaced equally in time will not cause the same constructive effect for $\Delta_p\ne 0$. 

To take advantage of a pulse train one must start by exciting the two-level system. Immediately after excitation is generated, the couple laser between states $|\frac{5}{2},\frac{3}{2}\rangle \rightarrow |\frac{3}{2},\frac{3}{2}\rangle$ can be introduced. During the  time interval  when $\Omega_c$ is on, the quantum beat is visible with an increased intensity due to the larger excitation in the sample. Then, only after the couple laser is turned off should the sample be further excited.

\section{Applied static magnetic field \label{2samples}}

\begin{figure}[h!]
\centering
\includegraphics[width=0.45\textwidth]{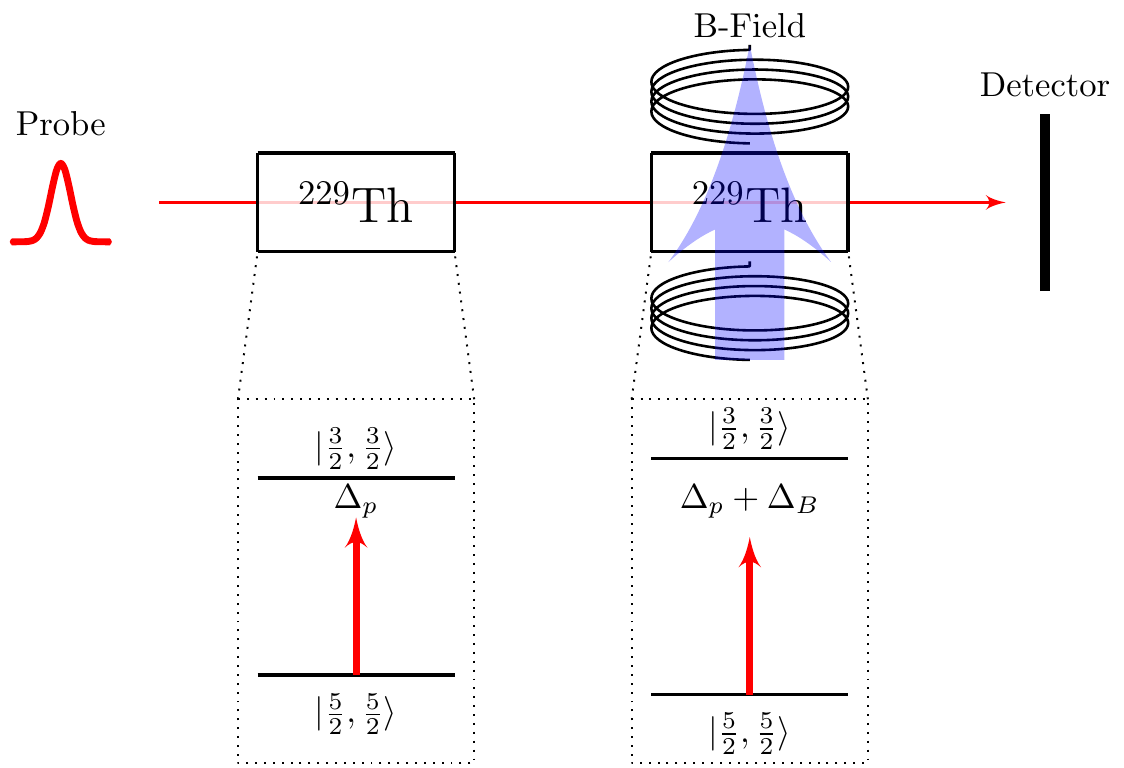}
\caption{Two $^{229}$Th:CaF$_{2}$ target setup. A left circularly polarized probe field drives the $\vert\frac{5}{2},\frac{5}{2}\rangle\leftrightarrow\vert\frac{3}{2},\frac{3}{2}\rangle$ isomeric transitions in both crystals. The detuning of the probe to the unperturbed resonance frequency is denoted by $\Delta_{p}$. $\Delta_{B}$ denotes the total Zeeman shift due to the external magnetic field B.}
\label{fig:2samples}
\end{figure}

\begin{figure}[h!]
\centering
\includegraphics[scale=1]{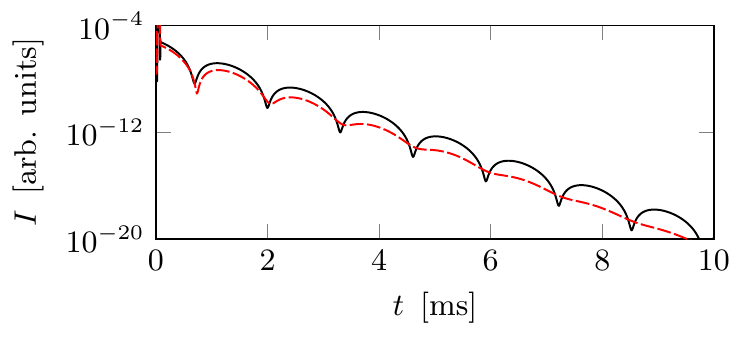}
\caption{Scattered intensity for the two crystal system with $B=10^{-4}$ T for  $\Delta = 0 \sim 10^6\Gamma$ (black solid line) and  $\Delta = 10^8 \Gamma$ (red dashed line). Excitation occurs for $\Delta < 10^{10} \Gamma$.}
\label{fig:mag1}
\end{figure}

As mentioned in Sec.~\ref{numres1}, quantum beating can be induced also without the use of a second laser. Instead, we can consider two crystals excited one after the other by the same laser pulse as illustrated in Fig.~\ref{fig:2samples}. By placing one of the two crystals in a static magnetic field the energy levels will be shifted by $\Delta_B = (m_g \mu_g+ m_e\mu_e)B/\hbar$, where $(\mu_g,\mu_e,B) = ( 0.45 \mu_N, -0.08 \mu_N, 10^{-4} T)$ and $\mu_N = 5.05 \times 10^{-27}$ J/T. The NFS spectrum is then the result of a pair of two-level systems which mimic the results of a three-level system. The resulting quantum beat shown in Fig.~\ref{fig:mag1} is due to the additional energy shift of the levels in the static magnetic field, resulting in a period $T_{QB} = 2\pi/\Delta_B$.

\begin{figure}[h!]
\centering
\includegraphics[scale=1]{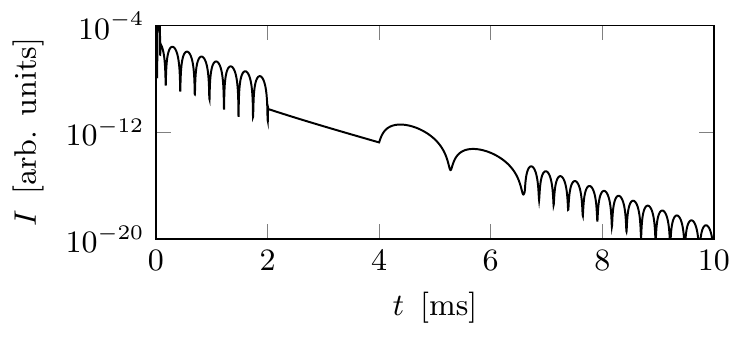}
\caption{Scattered intensity for the two crystal system considering $\Delta = 0$. The applied external magnetic field takes the values $B=5\times 10^{-4}$ T, $0$ T, $10^{-4}$ T, and again $5\times 10^{-4}$ T.}
\label{fig:mag2}
\end{figure}

Analogous to the role of the coupling laser in the three-level system, the static magnetic field can be turned off and on as well as adjusted in magnitude. This allows for switching between a two-level decay and a three-level quantum beat of variable frequency seen in Fig.~\ref{fig:mag2}, where the applied external magnetic field changes value at the start or end of a beat cycle corresponding to times $t_{switch} \approx (2,4,6.6)$ ms.

This setup is more advantageous than the three-level system outlined above because it requires only one tunable laser. Furthermore, when undergoing multipulsed excitation the oscillation frequencies of $\Re \{\Omega\}$ and $\Im\{\Omega\}$ are the same as that of the two-level  system $f_\Omega = \Delta/{2\pi}$ ($f_{\Omega^2} = \Delta/{\pi}$) and thus the intensity can be reliably changed by varying the frequency of probe pulses while the static field is on. This removes the complication of switching off the couple laser during excitation as would be required in the three-level case. 

As a side remark, a slightly modified level scheme becomes available when considering the $90^\circ$ interstitial configuration or when thorium is doped in a different crystal such as LiCaAlF$_{6}$. In these cases, due to the change in sign of the electric-field gradient, the lowest-energy level corresponds to $|\frac{5}{2}, \pm\frac{1}{2}\rangle$. Hence, in a cooled crystal this state will be initially populated, allowing for two $\Delta m=0$ transitions to be driven at the same energy. Applying a static magnetic field will split these two transitions in opposite directions due to the sign of $m$, which results in quantum beating without the use of a second crystal. Thus, in such crystals the two-crystal setup discussed here can be reduced to only a single crystal in a static magnetic field.

\section{Conclusions \label{Conclusions}}
Excitation schemes for both two- and three-level systems in $^{229}$Th:CF$_2$ crystals have been investigated theoretically. We have shown that the complex crystal system with 10 levels and three q axes can be understood when looking at simplified systems of two- and three-level systems of a single quantization axis. These systems can be selectively driven with polarized fields provided the detuning is known such that $E_{\Delta i}<E_p$. For such systems it has been shown that both the time delay and phase shift between excitation pulses can cause a change in intensity of the measured signal in NFS experiments. To reliably increase the signal intensity using multiple excitation pulses, the detuning of the driving laser to the transition of interest must be known. Once initial excitation is found in the two-level system, multipulsed excitation can be used to increase the intensity of the signal and determine laser detuning in the system. Additionally, a signature of excitation in the form of quantum beating can be created by ($i$) a second laser to couple the $|\frac{5}{2},\frac{3}{2}\rangle$ and $|\frac{3}{2},\frac{3}{2}\rangle$ levels or ($ii$) a second crystal in a static magnetic field. These findings are anticipating first coherent driving of the thorium nucleus with VUV sources, which have so far failed mainly due to our poor knowledge of the transition frequency. A more exact value for the latter remains a prerequisite for any attempt to directly excite with lasers the nuclear isomer.

\begin{acknowledgments}
B.S.N. and A.P. gratefully acknowledge funding by the EU FET-Open project, Grant No. 664732 ``NuClock". W.-T.L. is supported by the Taiwan Ministry of Science and Technology, Grant No. MOST 107-2112-M-008-007-MY3. W.-T.L. also acknowledges support from the National Center for Theoretical Sciences and Center for Quantum 
Technology, Hsinchu, Taiwan.
\end{acknowledgments}

\appendix
\section{Crystal formation and quantization axis \label{AppA}}

\begin{figure*}[htb!]
\centering
\includegraphics[scale=0.5]{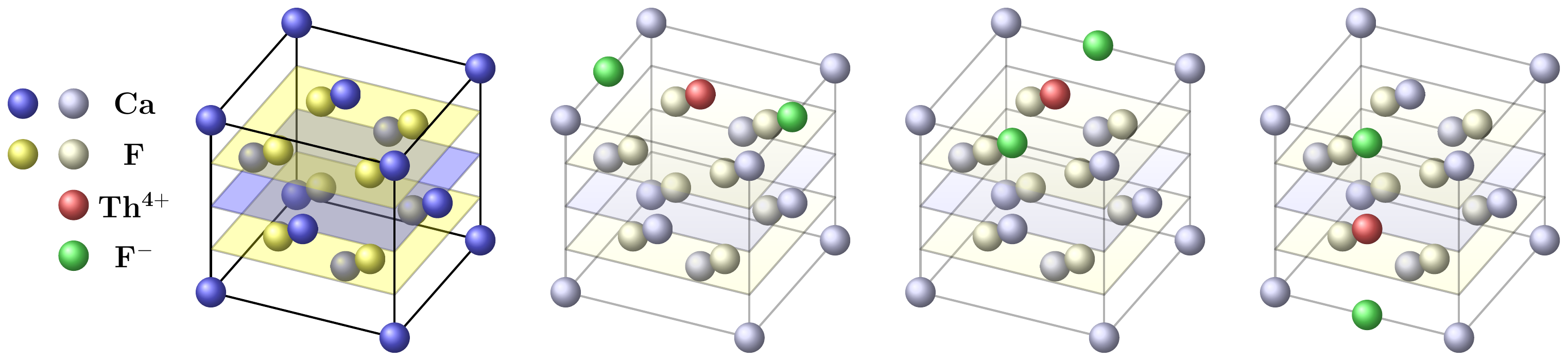}
\caption{Th:CaF$_2$ structure with $180^\circ$ fluoride interstitials showing the three possible rotations allowed in a bulk crystal.}
\label{fig:CaF2180}
\end{figure*}

The cubic lattice of CaF$_2$ is identical when rotated in the $xy$, $yz$, or $zx$ planes in increments of $90^\circ$.  As a result of this rotational symmetry, at low dopant densities where dopant sites do not interact there are three possible orientations of the $180 ^\circ$ F-Th-F bond, Fig. \ref{fig:CaF2180}. All three orientations are populated by $1/3$ of the total dopant density $N$. The quantization axis is fixed by the electric-field gradient at the location of the Th nuclei, which in this case is along the F-Th-F bond. Hence the three $q$ axes are mutually perpendicular and can be aligned with the laboratory-frame axes by rotating the crystal and viewing the resulting spectrum.

The energy splitting of all three orientations are the same; however, the angular momentum projections are along the mutually perpendicular $q$-axis directions. To understand what transitions will be driven we transform the driving field polarization vector into the reference frame of the dopant nuclei. $\Delta m = 0$, i.e., $\pi$ transitions are driven by polarization along, and $\Delta m = +(-)1$, i.e., $\sigma^{+(-)}$ require right(left)-circular polarization around the the subsystems $q$ axis. The resultant spectrum is then a combination of the excitation of all three orientations and all driven states. 

\begin{figure}[htb!]
\centering
\includegraphics[scale=1]{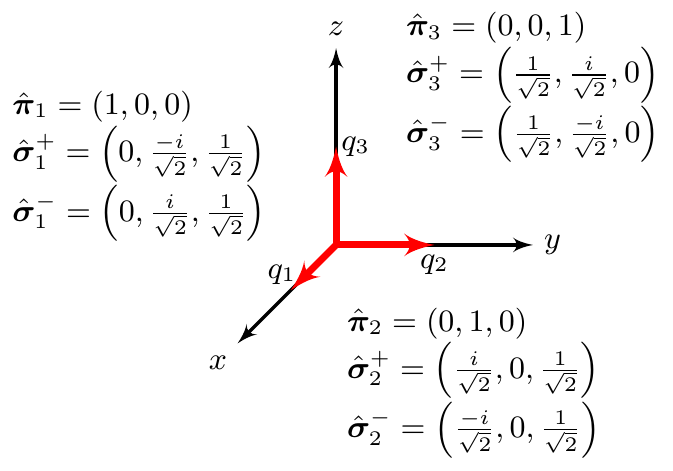}
\caption{Polarization vectors of fields used to drive the $(\pi,\sigma^+, \sigma^-)$ transitions relative to the laboratory frame $(x,y,z)$.}
\label{fig:qaxisproj}
\end{figure}
Looking at Fig.~\ref{fig:thorium} there are 10 states and 12 possible transitions, four of each $(\pi,\sigma^+,\sigma^-)$, for each of the three $q$ axes $q=(x,y,z)$. By introducing a polarized field, for example $\sigma^-_y$ (left-circular polarization in the direction $y$ in the laboratory frame), only $\sigma^-$ transitions can be driven for dopant nuclei with $q$ axis along $y$ ($1/3$ of the population). The same $\sigma^-_y$ polarization can be broken down into components which satisfy the selection rules for driving all transition in the other two orientations $(\pi_{x/z},\sigma^+_{x/z}, \sigma^-_{x/z})$ (see Fig.~\ref{fig:qaxisproj}) albeit with lower intensity. 

Clearly this situation appears complicated; however, excitation by one and two fields can be simplified to resemble the two- and three-level schemes, respectively shown in Fig. \ref{fig:thorium}. The key to this simplification becomes obvious only when you consider the relative energy scales of the level widths and their separation, as well as that of the laser field used for excitation. Introducing some notation, let the linewidth of the $^{229m}$Th transition in energy be $E_\gamma$, the energy spread of a laser pulse with half width in time $T$ be $E_p = \hbar/T$, and let the detuning of the laser to the $i^{th}$ transition be $E_{\Delta i}=\hbar \Delta_i$. $E_\gamma\approx 4\times 10^{-20}$ eV, which is so narrow that it can be considered an exact energy on the scale of the energy spread of the laser pulses used as well as on the scale of the quadrupole level splitting which is $E_Q=\mathcal{O}(10^{-6})$ eV. If $E_{\Delta i}>E_p$ there is negligible excitation of the $i^{th}$ transition by the pulse when compared to transitions of smaller detuning. Based on numerical limitations we use pulse widths on the order of $\mu$s making the energy spread $E_p= \mathcal{O}(10^{-10})$ eV. Therefore, when considering driving a transition close to resonance, the quadrupole splitting rules out the possibility for the same pulse driving more than two transitions, where the two transitions are degenerate in energy differing only in the sign of the momentum projections $\pm \Delta m$. As a result, we have $E_\gamma \ll E_{\Delta_i}<E_p<E_Q$ and transitions away from resonance and corresponding states can be safely neglected during calculation. 

\begin{figure}[htb!]
\centering
\includegraphics[scale=1]{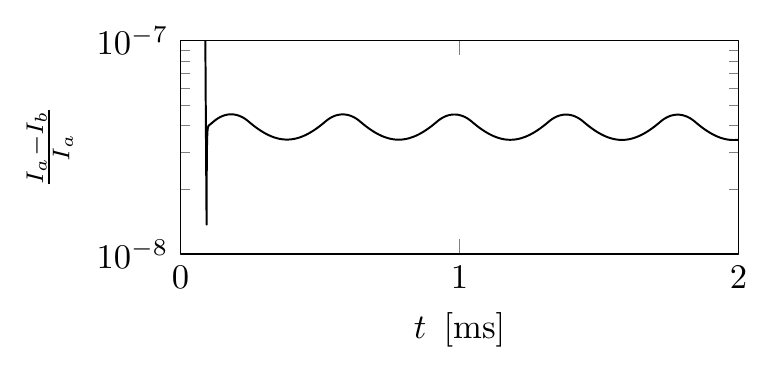}
\caption{  Relative difference in spectrum intensity $(I_a-I_b)/I_a$ considering (a) two-state system, 1$q$-axis, $N/6$ initial population of single ground state $|\frac{5}{2},\frac{5}{2} \rangle$, and (b) 10-state system, 1$q$-axis, initial population $N/3$ split equally among two lowest states $|\frac{5}{2},\pm\frac{5}{2} \rangle$. The calculations were performed considering $\Delta = 0$ for the transition $|\frac{5}{2},\pm \frac{5}{2} \rangle \rightarrow |\frac{3}{2},\pm\frac{3}{2} \rangle$.}
\label{fig:1qaxiscomp2v10}
\end{figure}
This can be seen when comparing calculations of 10-state and two-state systems of a single $q$ axis. Continuing the earlier example, a $\sigma^-_y$ field is used in resonance with $|\frac{5}{2},\pm \frac{5}{2} \rangle \rightarrow |\frac{3}{2},\pm\frac{3}{2} \rangle$ transition. In this case all other transitions satisfy $E_{\Delta i}>E_p$ and thus can be neglected. Figure \ref{fig:1qaxiscomp2v10} shows the difference between the two spectrum divided by the total intensity at each point in time. The scale of this difference reveals that removing the additional levels for which $E_{\Delta i}>E_p$ has a negligible effect on the spectrum, making both calculations effectively  identical. 

\begin{figure}[htb!]
\centering
\includegraphics[scale=1]{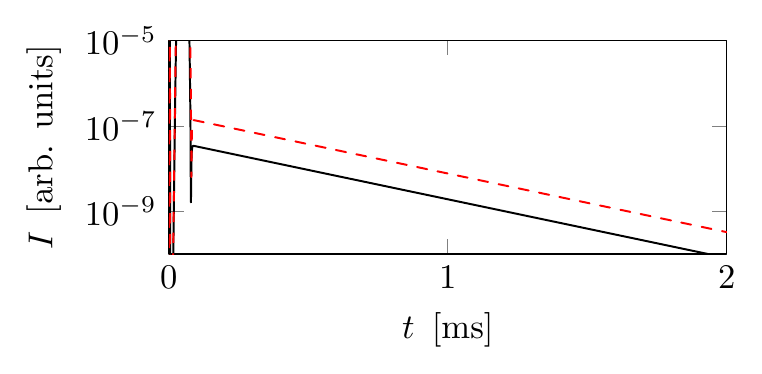}
\caption{ Scattered intensity for (black solid line) two-state system, 1$q$-axis, initial population $N/6$ in $|\frac{5}{2},\frac{5}{2} \rangle$ and (red dashed line) four-state systems, three q axes, $N$ initial population split equally among two lowest states $|\frac{5}{2},\pm\frac{5}{2} \rangle$ with $1/3$ in each possible $q$-axis orientation. The calculations were performed considering $\Delta = 0$ for the transition $|\frac{5}{2},\pm \frac{5}{2} \rangle \rightarrow |\frac{3}{2},\pm\frac{3}{2} \rangle$. }
\label{fig:qaxiscomp2_1qv4_3q}
\end{figure}

At this point it is clear that the calculation can be simplified by removing levels far out of resonance. Now let us consider the effect of multiple quantization axes in a single sample. We recall  that $\sigma^-_y$ field has components along $(\pi_{x/z},\sigma^+_{x/z}, \sigma^-_{x/z})$, i.e., fractions of the field have correct polarization to drive all transitions in the other two $q$-axis orientations. However, the frequency of the field is only in resonance with the $\sigma^{\pm}$ corresponding to $|\frac{5}{2},\pm \frac{5}{2} \rangle \rightarrow |\frac{3}{2},\pm\frac{3}{2} \rangle$. Hence we neglect all other levels which reduces the 10 states to four states. Figure \ref{fig:qaxiscomp2_1qv4_3q} compares the four-state calculation with three q axes to that of the simple two-state system of only one q axis. 

In this case the calculations compare in their functional behavior yet yield differing intensities. If the intensity of the signal at time $t$ for the two-state calculation with one q axis is $I^2_1(t)$ and four state calculation with three q axes is $I^4_3(t)$, respectively, then $I^4_3(t)\approx 4I^2_1(t)$. This is because the intensity of the signal is proportional to the square of the population making the transition, $I\propto \rho^2$. The intensity of the three q axis system is greater because $\sigma^-_y$ can drive more of the population to the excited state due to its projection on the other two $q$ axes, which were neglected in the simplified two-state calculation. Vector projection gives fields of half amplitude driving $\sigma^{+/-}_{x/z}$, which result in a contribution of $\approx\frac{1}{2}I^2_1(t)$ from each of the $x$ and $z$ $q$-axis populations. These then add in superposition with matching polarizations to give the final result,
\begin{eqnarray}
I^4_3 &\approx& \left(A_x +\frac{A_x}{2}+\frac{A_x}{2}\right)^2+\left(A_z +\frac{A_z}{2}+\frac{A_z}{2}\right)^2 \nonumber \\
&=& 4 (A_x^2+A_z^2) = 4 I_1^2.
\end{eqnarray}

The intensity differences written here correspond to exactly a combination of independent systems. Our numerical calculations show a slight deviation from this independent system treatment due to multiple scattering between $q$-axis subsystems. This difference is on the order of a few percent which can easily be neglected when looking at the main effects involved.

\section{Lowest energy configuration: $\eta\ne 0$ \label{AppB}}
The lowest-energy dopant configuration shown in \cite{Dessovic_2014} requires fluoride interstitials making a $90^\circ$ angle with the dopant thorium, which takes the position of a Ca shown in Fig. \ref{fig:CaF290180_2}(a).  The same symmetry applies where the crystal can be rotated in the $xy$, $yz$, or $zx$ planes in increments of $90^\circ$; only in this case due to the angled F-Th-F bond there are 12 orientations. This is not a problem as the electric-field gradient runs perpendicular to the plane defined by the angled F-Th-F bond making the 12 orientations fourfold degenerate. As such there are only three orientations of the quantization axis, just like before. 

\begin{figure}[h!]
\centering
\includegraphics[scale=1]{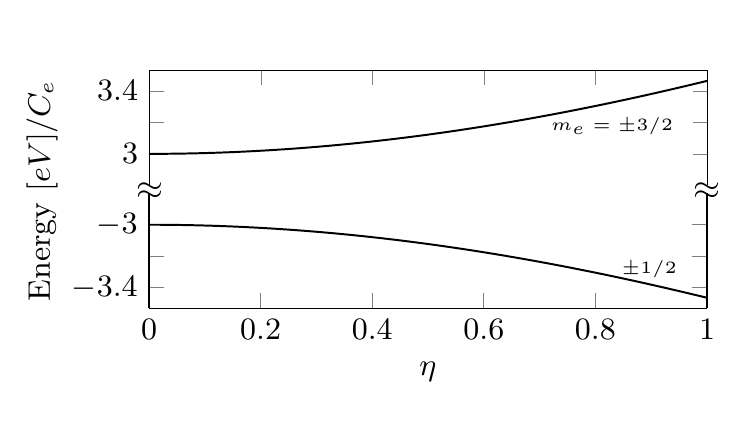}\\
\caption{Energy eigenvalues of $\hat{H}_{E2}$ for $I=I_e=3/2$ as a function of the asymmetry parameter $\eta$. Labeled with the spin projection $m_e$ with respect to the largest field component.}
\label{fig:E3/2}
\end{figure}

\begin{figure}[h!]
\centering
\includegraphics[scale=1]{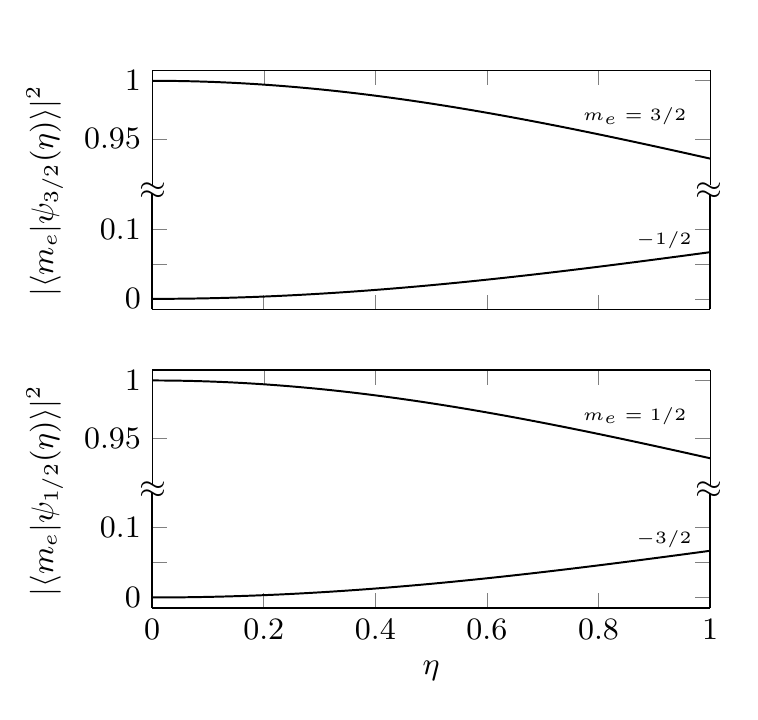}
\caption{Squared projection of state vector $|\frac{3}{2},\psi_m(\eta)\rangle$ on overlapping eigenstates of angular momentum as a function of the asymmetry parameter $\eta$. }
\label{fig:state3/2}
\end{figure}

\begin{figure}[h!]
\centering
\includegraphics[scale=1]{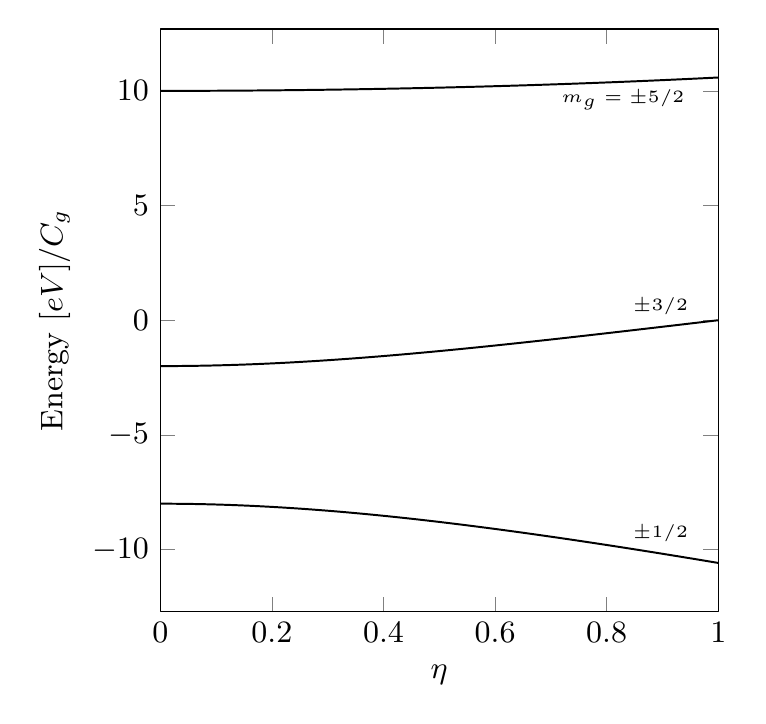}\\
\caption{Energy eigenvalues of $\hat{H}_{E2}$ for $I=I_g=5/2$ as a function of the asymmetry parameter $\eta$. Labeled with the spin projection $m_g$ with respect to the largest field component.}
\label{fig:E5/2}
\end{figure}

\begin{figure}[h!]
\centering
\includegraphics[scale=1]{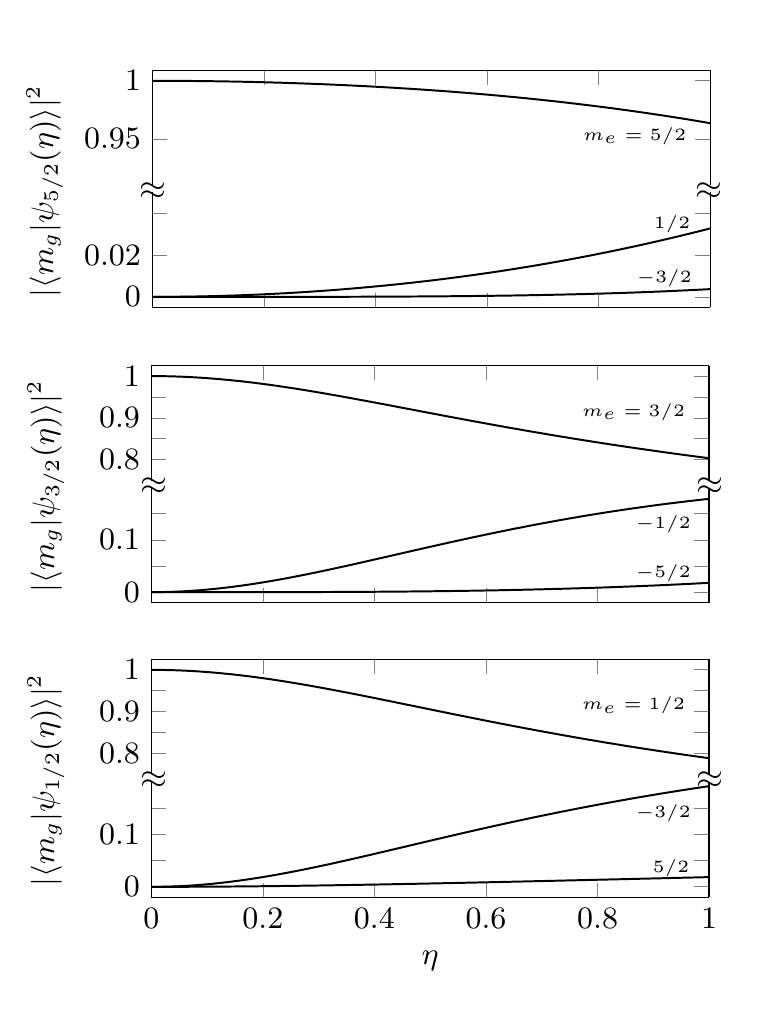}
\caption{Squared projection of state vector $|\frac{5}{2},\psi_m(\eta)\rangle$ on overlapping eigenstates of angular momentum as a function of the asymmetry parameter $\eta$. }
\label{fig:state5/2}
\end{figure}

In this dopant orientation $V_{zz}=223$ V\AA$^{-2}$ \cite{Dessovic_2014}, and the change in sign of the electric-field gradient reverses the ordering of the split levels in terms of energy as compared to the $180^\circ$ case shown in Fig.~\ref{fig:thorium}. Furthermore, the  $90^\circ$ configuration has a nonzero asymmetry parameter  $\eta=0.48$. In this case, the angular momentum projection $m$ is no longer a good quantum number to define the nuclear states. $0\le \eta \le 1$ and the larger $\eta$ becomes, the more mixing there is between sublevels. The Hamiltonian \eqref{eqn:He2} can be diagonalized to find the eigenvalues and eigenvectors, which are shown here for the ground $I_g=5/2$ and excited (isomer) state $I_e=3/2$ \cite{Matthias1962, Collins1967, Suits2006, Cho2016}. Figures \ref{fig:E3/2} and \ref{fig:E5/2} show the energy of the split ground and excited states where $C_{g/e} = eQ_{g/e} V_{zz} / (4I_{g/e}(2I_{g/e}-1))$. Figures \ref{fig:state3/2} and \ref{fig:state5/2} show the squared projection of the state vectors with the eigenstates of nuclear spin angular momentum.  
The energy splitting is still $\mathcal{O}(10^{-6})$ eV. Therefore, once again when driving a transition close to resonance, the quadrupole splitting rules out the possibility for the same pulse driving more than two transitions, where the two transitions are degenerate in energy differing only by the sign of the states' $m$ projections.  

What is different in this case is that the angular momentum selection rules can be satisfied in more than one way for a given transition. Each state can now be defined as a superposition of angular momentum eigenstates where the probability to be in a given eigenstate is given by $|\langle I, m'|I,\psi_m(\eta)\rangle|^2$, where $|I,m\rangle$ is an eigenstate of angular momentum with projection $m$ and $|I,\psi_{m'}(\eta)\rangle = \sum_m a_m(\eta) |I,m\rangle$ is the state vector with the largest contributing component coming from state $|I,m'\rangle$. For example, consider the transition from $|\frac{5}{2},\xi_{1/2}(\eta)\rangle \rightarrow |\frac{3}{2},\phi_{1/2}(\eta)\rangle$. If the laser has the correct energy within the energy width of the pulse, then only the polarization determines whether the transition will be driven. Expanding the state vectors we see
\begin{eqnarray}
|\tfrac{5}{2},\xi_{1/2}(\eta)\rangle &=& a_1|\tfrac{5}{2},\tfrac{1}{2}\rangle +a_2|\tfrac{5}{2},-\tfrac{3}{2}\rangle+ a_3|\tfrac{5}{2},\tfrac{5}{2}\rangle, \nonumber \\
|\tfrac{3}{2},\phi_{1/2}(\eta)\rangle &=& b_1|\tfrac{3}{2},\tfrac{1}{2}\rangle +b_1|\tfrac{3}{2},-\tfrac{3}{2}\rangle,
\end{eqnarray}
and hence only $\Delta m=0$ transitions are possible. To understand the relative strength of the transitions we calculate the square of the overlap integral $|\langle I_g,m_g|\langle J,M|I_e,m_e\rangle|^2$ \cite{Collins1967, Rohlsberger2004}, where the excited state $|I_e,m_e\rangle$ is connected
to the ground state $|I_g,m_g\rangle$ and the emitted photon $|J,M\rangle$ via the Clebsch-Gordan coefficients
\begin{eqnarray}
|I_e,m_e\rangle = \sum_{M} \langle I_g,J,m_g,M|I_e,m_e\rangle |J,M\rangle |I_g,m_g\rangle.  \nonumber \\
\end{eqnarray}

Hence for the above transition,
\begin{eqnarray}
\langle \tfrac{5}{2},\xi_{1/2}(\eta)|\langle 1, 0|\tfrac{3}{2},\phi_{1/2}(\eta)\rangle &=&a_1b_1\langle\tfrac{5}{2},1,\tfrac{1}{2},0|\tfrac{3}{2}, \tfrac{1}{2}\rangle \nonumber\\
&+& a_2b_2\langle\tfrac{5}{2},1,-\tfrac{3}{2},0|\tfrac{3}{2}, -\tfrac{3}{2}\rangle \nonumber\\
&=& -a_1b_1\sqrt{\tfrac{2}{5}} - a_2b_2\tfrac{2}{\sqrt{15}}, \nonumber
\end{eqnarray}
where the coefficients for $\eta=0.48$ are $(a_1,a_2,a_3,b_1,b_2) = (0.95,0.29,0.08,0.99,0.13)$ and correspond to the square root of values taken from Figs. \ref{fig:state3/2} and \ref{fig:state5/2}. Comparing the intensity to a transition between pure eigenstates of angular momentum (i.e., the unmixed case of $\eta=0$) results in
\begin{eqnarray}
\frac{\left|\langle \tfrac{5}{2},\xi_{1/2}(0.48)|\langle 1, 0|\tfrac{3}{2},\phi_{1/2}(0.48)\rangle\right|^2}{\left|\langle \tfrac{5}{2},\frac{1}{2}|\langle 1, 0|\tfrac{3}{2},\frac{1}{2}\rangle\right|^2} &\approx & 0.98.
\end{eqnarray}
Hence the intensity of this transition is lowered by $\approx 2\%$ due to mixing of states. In general, for our case $|\langle I, m|I,\psi_m(0.48)\rangle|^2 > 0.9$ for all $I$ and $m$. As a result the intensity of the signal will drop in relation to the smaller population undergoing the transition. Clearly, this is negligible for our purposes. As such all results shown above apply in this case with minor intensity corrections. 

The one exception is the two-crystal setup discussed in Sec. \ref{2samples}, which can be reduced for the case of $90^\circ$ dopant orientation to only a single crystal in a static magnetic field. Due to the change in sign of the electric-field gradient the lowest-energy level corresponds to $|\frac{5}{2}, \pm\frac{1}{2}\rangle$. Hence, in a cooled crystal, this state will be initially populated allowing for two $\Delta m=0$ transitions to be driven at the same energy. Applying a static magnetic field will split these two transitions in opposite directions due to the sign of $m$, which results in quantum beating without the use of a second crystal, similar to the case of the LiCaAlF$_6$ crystal discussed in Sec. \ref{2samples}.

\vfill
\newpage

\newpage	
\bibliographystyle{unsrt}
\bibliography{references}

\end{document}